\documentclass[aps,prd,twocolumn,showpacs]{revtex4}
\usepackage{epsfig}

\def\to{\rightarrow}

\def\bi{\begin{itemize}}
\def\ei{\end{itemize}}
\def\be{\begin{equation}}
\def\ee{\end{equation}}
\def\bea{\begin{eqnarray}}
\def\eea{\end{eqnarray}}

\def\tb{\tilde b}

\def\ttau{\tilde \tau}

\usepackage{colordvi}

\begin{document}

\title{Discovering Bottom Squark Co-annihilation at ILC}

\author{Alexander Belyaev}\affiliation{School of Physics \& Astronomy, University of Southampton,\\ Highfield, Southampton SO17 1BJ, UK}\affiliation{Particle Physics Department, Rutherford Appleton Laboratory, \\Chilton, Didcot, Oxon OX11 0QX, UK}
\author{Tom\'{a}\v{s} La\v{s}tovi\v{c}ka}\affiliation{University of Oxford,\\ Denys Wilkinson Building, Keble Road,\\ Oxford OX1 3RH, UK}
\author{Andrei Nomerotski}\affiliation{University of Oxford,\\ Denys Wilkinson Building, Keble Road,\\ Oxford OX1 3RH, UK}
\author{Gordana La\v{s}tovi\v{c}ka-Medin}\affiliation{University of Montenegro,\\ Cetinjska bb,\\ 81 000 Podgorica, Montenegro}

\begin{abstract} We study the potential of International Linear Collider (ILC)  at
\mbox{$\sqrt{s} = 500$\,GeV} to probe new dark matter motivated scenario where the
bottom squark (sbottom)  is the next to lightest supersymmetric particle. For this
scenario, which is virtually impossible  for the LHC to test, the ILC has a potential
to cover a large fraction of the parameter space. The challenge is due to a very low
energy of jets, below 20-30\,GeV, which pushes the jet clustering and flavour tagging
algorithms to their limits. The process of sbottom pair production was studied within
the SiD detector concept. We demonstrate that ILC offers a unique opportunity to test
the SUSY parameter space motivated by the sbottom neutralino co-annihilation scenario
in cases when the sbottom production is kinematically accessible. The study was done
with the full SiD simulation and reconstruction chain including all Standard Model
and beam backgrounds. \end{abstract}


\maketitle


\section{Introduction}

Among the best candidates for  a theory beyond the Standard Model (SM), Supersymmetry
(SUSY)~\cite{Golfand:1971iw,Ramond:1971gb, Neveu:1971iv,
Volkov:1973ix,Wess:1974tw,Haag:1974qh} remains a very compelling theory even after 30 years
without an experimental confirmation. SUSY is very attractive because it successfully
solves principal theoretical and experimental problems of the SM. Supersymmetric theories
provide a natural solution to the gauge hierarchy problem of the SM and incorporate the
unification of gauge coupling constants. Furthermore, the lightest supersymmetric particle
(LSP) is stable if the R-parity is conserved and can serve as a good cold dark matter (CDM)
candidate. Besides, SUSY  has all ingredients to provide a solution to the baryogenesis
problem via the intermediate scale leptogenesis~\cite{Fukugita:1986hr} or via the
electroweak baryogenesis~\cite{Giudice:1992hh,Carena:1997gx}.

The very existence of CDM is a crucial argument in favour of SUSY, and, at the same time,
an important SUSY constraint. The most direct evidence for CDM in the Universe comes from
observations of galactic rotation curves. Binding of galaxies in clusters, matching
observations of large scale structure with simulations, gravitational microlensing,
baryonic density of the Universe as determined by Big Bang nucleosynthesis, observations of
supernovae in distant galaxies, as well as measurements of anisotropies in the cosmic
microwave background radiation (CMB) can also be considered as strong confirmations of CDM
(for reviews see e.g. \cite{Freedman:2003ys,cdm-nanopoulos}). In particular, the analysis
of the Wilkinson Microwave Anisotropy Probe (WMAP) and galaxy survey data  puts the most
stringent constraint on the ratio of the dark matter density to the critical
density~\cite{Spergel:2006hy}:
\begin{equation}
\label{eq:wmap}
\Omega_{CDM}h^2 =0.111^{+0.011}_{-0.015} \ ({\rm at\  95\% CL}),
\end{equation}
where $h = 0.74\pm 0.03$ is the normalized Hubble constant.

In most of the parameter space of  SUSY models, the value of $\Omega_{CDM}h^2$ is well
above the WMAP bound. In particular, in case of well explored minimal supergravity (mSUGRA)
model~\cite{Chamseddine:1982jx,Barbieri:1982eh,Ohta:1982wn,Hall:1983iz}, which is defined
by universal soft SUSY breaking scalar masses ($m_0$), gaugino masses ($m_{1/2}$) and
A-terms ($A_0$) at GUT scale, the CDM is the lightest neutralino $\tilde{\chi}_1^0$.
Neutralino should annihilate or co-annihilate with other SUSY particles at the Early
Universe time intensively enough to lower down
$\Omega_{CDM}h^2=\Omega_{\tilde{\chi}_1^0}h^2$ to experimentally acceptable level. This
happens in the following special regions of SUSY parameter space:

\begin{enumerate}

\item The bulk annihilation region at low values of $m_0$ and $m_{1/2}$, where neutralino
pair annihilation occurs at a large rate via $t$-channel slepton exchange.

\item The stau co-annihilation region at low $m_0$ where $m_{\tilde{\chi}_1^0}\simeq
m_{\ttau_1}$ so that $\tilde{\chi}_1^0$s may co-annihilate with $\ttau_1$s in the early
universe~\cite{Ellis:1998kh,Ellis:2000we}.

\item The hyperbolic branch/focus point (HB/FP)
region~\cite{Chan:1997bi,Feng:1999zg,Baer:1995va,Baer:1995nq} at large $m_0$ near the
boundary of the Radiative Electroweak Symmetry Breaking (REWSB) excluded region where the
superpotential Higgsino mass term $|\mu |$ becomes small and the neutralinos have a
significant higgsino component, facilitating their annihilations into $WW$ and $ZZ$ pairs.

\item The $A$-annihilation funnel, which occurs at very large values of
\mbox{$\tan\beta\sim
45-60$}~\cite{Drees:1992am,Baer:1995nc,Baer:1997ai,Baer:2000jj,Ellis:2001ms,Roszkowski:2001sb}.
In this case, one has $m_A\sim 2m_{\tilde{\chi}_1^0}$. An exact equality in the mass
relation is not necessary, since the $A$ width can be quite large ($\Gamma_A\sim
10-50$\,GeV); one can thus achieve a large $\tilde{\chi}_1^0\tilde{\chi}_1^0\to A\to
f\bar{f}$ annihilation cross section even if $2m_{\tilde{\chi}_1^0}$ is several values of
width away from resonance. The heavy scalar Higgs $H$ also contributes to the annihilation
cross section.

\end{enumerate}

In addition, for a particularly large $A_0$ values there exists a region of
neutralino top-squark (stop) co-annihilation and, at low $m_{1/2}$ values, a light
Higgs $h$ annihilation funnel region.

In all these regions the processes of neutralino \mbox{(co-)annihilation} have a high
enough rate to suppress the value of $\Omega_{\tilde{\chi}_1^0}h^2$ to the
experimentally acceptable level.

One should note that each particular DM motivated region of SUSY parameter
space defines a specific collider phenomenology.  It is also worth to
mention that some of those regions are problematic for the LHC and the ILC
would play a crucial role in exploring them or even in discovering
Supersymmetry in these regions. The representative example of such a region
is the HB/FP region which, as it was shown, can be efficiently covered only
by the 1\,TeV ILC collider\cite{Baer:2003wx,Baer:2003jb,Baer:2004qq}.

In this paper we study the sbottom-neutralino co-annihilation (SBC)
scenario and its ILC phenomenology. In the SBC scenario, as follows below, the typical
sbottom-neutralino relative mass difference $\delta m_{\tilde{b}\tilde{\chi}_1^0}
= (m_{\tilde{b}}-m_{\tilde{\chi}_1^0})/m_{\tilde{\chi}_1^0}$ is below  $10\%$.
This parameter space is inaccessible neither at Tevatron nor at the LHC.
It is interesting to notice that Tevatron would be in a better position than the LHC
to study light sbottom scenario which requires very low $p_T$ $b$-jet
threshold and would be sensitive to kinematically allowed parameter space
with $\delta m_{\tilde{b}\tilde{\chi}_1^0}\gtrsim 40\%$~\cite{Baer:1998kt,Demina:1999ty}.
Therefore the SBC scenario which we study here is new for the collider phenomenology.
Also, as we show below, the SBC scenario has not been studied in detail from the
theoretical point of view.

The rest of the paper is organized as follows. In Section II we discuss in detail the
SBC scenario and, in particular, the GUT scale boundary conditions providing SBC at
the Electroweak Scale. In Section III we present the signal and background studies at
the detector level and work out a strategy for observation of the signal. Conclusions
are drawn in Section IV.

\section{Sbottom co-annihilation scenario}

As we mentioned above, one of the important mechanisms of neutralino relic density
suppression occurs when the next-to-lightest supersymmetric particle (NLSP), is close in
mass to the neutralino. In this case, the neutralino relic density is not only suppressed
by the neutralino-neutralino annihilation, but also by co-annihilation with the NLSP.

In the traditionally well-explored mSUGRA model only stau or stop co-anninhilation can take
place to suppress the CDM relic density down to the experimentally allowed region.

On the other hand in the various models motivated by SUSY GUTs one can expect different
kinds of deviation from universal boundary conditions which would predict qualitatively
different phenomenology still consistent with the present experimental data (see e.g.
\cite{Baer:2008ih} and references therein).

In such models some new co-annihilation scenarios can take place. In particular, in
\cite{Pallis:2003aw,Profumo:2003ema}  it was noticed that under certain relation of scalar masses
the sbottom-neutralino relative mass difference $\delta m_{\tilde{b}\tilde{\chi}_1^0}
= (m_{\tilde{b}}-m_{\tilde{\chi}_1^0})/m_{\tilde{\chi}_1^0}$ can become below of
about $10\%$ and provide effective SBC suppression of the relic density down to the
experimentally acceptable level. The papers~\cite{Pallis:2003aw,Profumo:2003ema} concluded that in
case of minimal sfermion non-universality (mSFNU) defined by $(\,m_{10}$, $ m_{5}$,
$A_0$, $m_{1/2}$, $\tan\beta,sign(\mu))\,$ parameter space the SBC scenario could be
realised, where the left-right squark and sfermion mass parameters at the GUT scale
extend the universal $m_0$ mSUGRA parameter.

$$m_5^2 =m_L^2=m_D^2$$
$$m_{10}^2=m_Q^2=m_U^2=m_E^2$$

It was found that the ratio $$K=m_5/m_{10}$$ below of about 0.5 could provide a SBC scenario.

In this section we take a closer look at the results of \cite{Pallis:2003aw,Profumo:2003ema} where
conditions for SBC were derived using simplistic one-loop  Renormalisation Group
Equations (RGEs). In our paper we use mass spectra calculation from ISAJET v7.79
package~\cite{Paige:2003mg} where the complete two-loop RGEs for the gauge couplings,
Yukawa couplings and soft breaking terms are implemented. Another reason for checking
of the SBC scenario was applying a different approach in comparison with
\cite{Pallis:2003aw,Profumo:2003ema}, where the {\it exact} $b-\tau$ unification was imposed at the
GUT scale and RGEs were run just once from this  scale. In the ISAJET the RGEs are
solved iteratively, running from the weak scale to the high scale and back, while
Yukawa couplings are not fixed at the GUT scale. After each iteration the SUSY masses
are recalculated, and the renormalization group improved one-loop corrected Higgs
potential is calculated and minimized. This way of mass calculation is widely
accepted to be the most accurate one.

We have performed an exhaustive scan of the SUSY parameter space defined by

\begin{eqnarray}
0\leq m_{1/2} \leq 2 \textrm{\,TeV}, & |A_0|<3 \mbox{\,TeV}, &  5<\tan \beta<50  \nonumber \\
0<m_{5} \leq 5 \textrm{\,TeV}        &&  \nonumber \\
0< m_Q<5\mbox{\,TeV}                 &&  \nonumber \\
0< m_U=m_E<5\mbox{\,TeV}
\label{eq:scan}
\end{eqnarray}

where we went beyond the minimal sfermion non-universality (mSFNU) scenario,  defined in
\cite{Pallis:2003aw,Profumo:2003ema},
and split $m_{10}=m_Q=m_U=m_E$ into two parameters $m_Q$ and $m_U=m_E$ while keeping
$m_{5}=m_L=m_D$ as defined previously.

\begin{figure*}[htbp]
\includegraphics[width=0.48\linewidth]{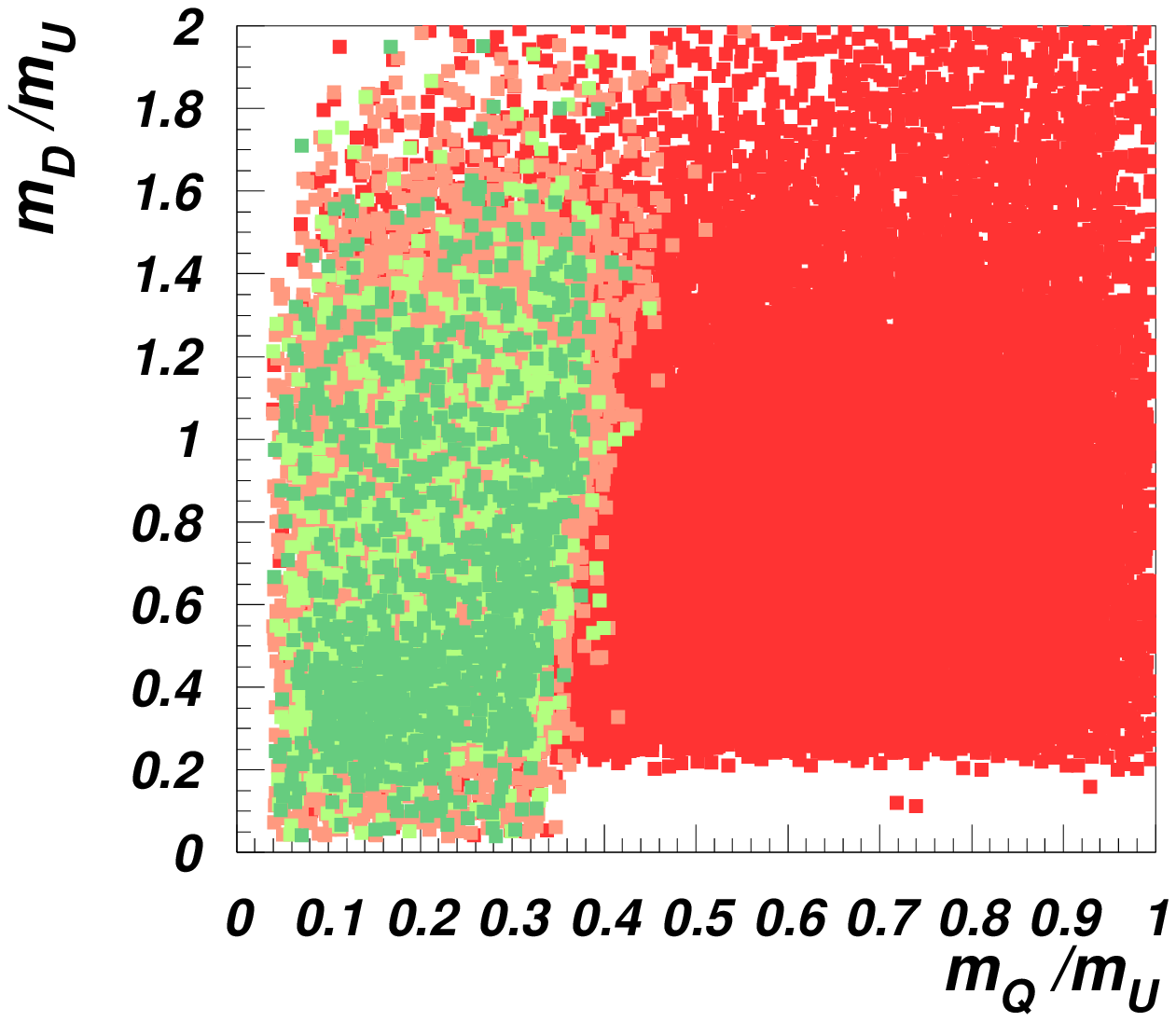}
\includegraphics[width=0.48\linewidth]{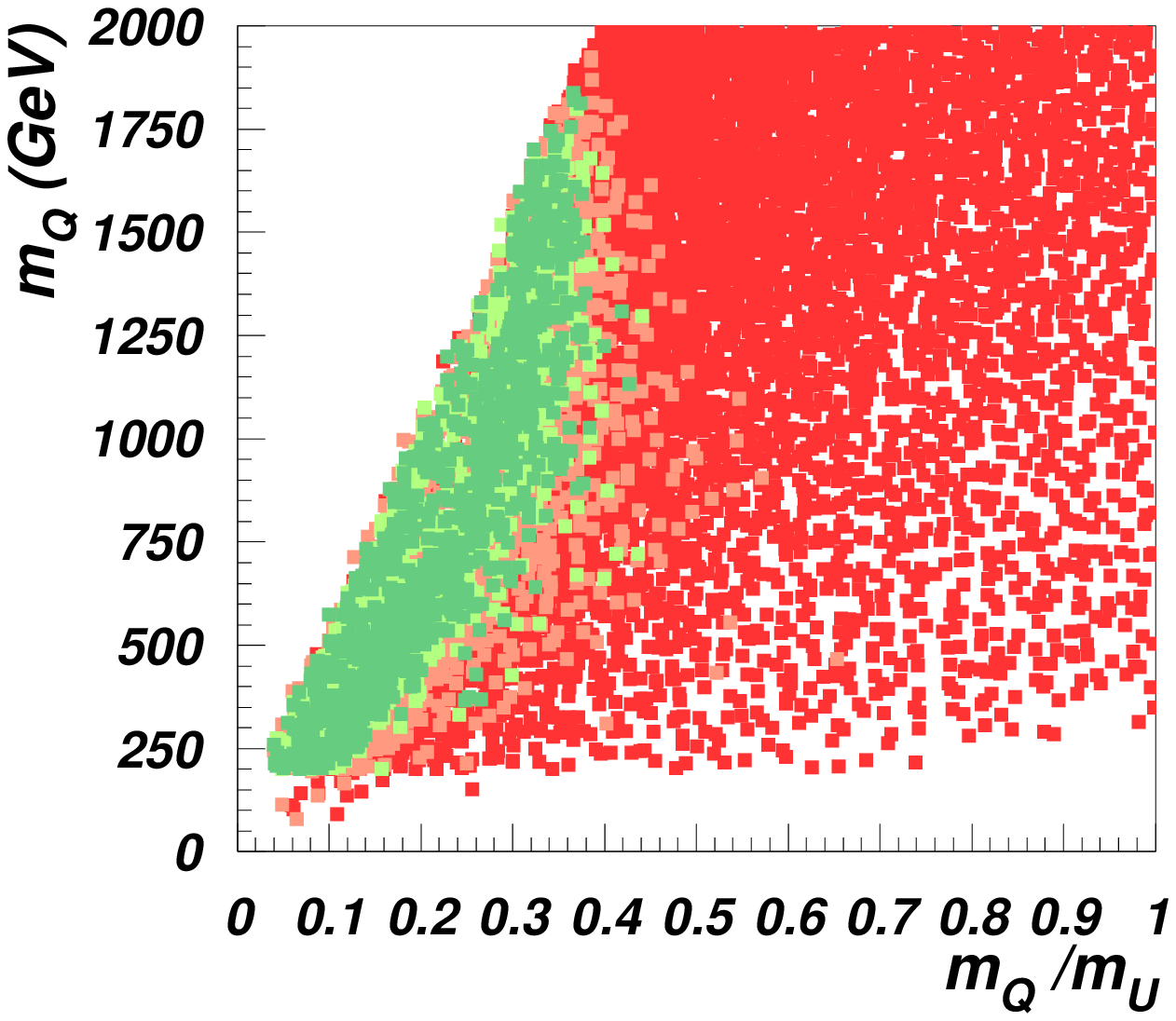}
\caption{\label{fig:scan1}
The results of the scan in SUSY parameter space defined by (\ref{eq:scan}) shown in
($m_D/m_U$-$m_Q/m_U$) (left) and ($m_Q$-$m_Q/m_U$) (right) planes respectively with the
following color code for $\delta m_{\tilde{b}\tilde{\chi}_1^0} =
(m_{\tilde{b}}-m_{\tilde{\chi}_1^0})/m_{\tilde{\chi}_1^0}$:\\
a) $1.0 < \delta m_{\tilde{b}\tilde{\chi}_1^0}$~(red/dark),\\
b) $0.2 < \delta m_{\tilde{b}\tilde{\chi}_1^0} < 1.0$~(pink/grey),\\
c) $0.1 < \delta m_{\tilde{b}\tilde{\chi}_1^0} < 0.2$~(light-green/very light grey),\\
d) $\delta m_{\tilde{b}\tilde{\chi}_1^0} < 0.1$~(dark-green/light grey).}
\end{figure*}

Our first results are shown in Figure~\ref{fig:scan1}, which presents the results of the scan in the SUSY parameter space defined by (\ref{eq:scan}) shown in $m_D/m_U$ vs. $m_Q/m_U$ (left) and $m_Q$ vs. $m_Q/m_U$ (right) planes respectively with the following color code for
$\delta m_{\tilde{b}\tilde{\chi}_1^0} = (m_{\tilde{b}}-m_{\tilde{\chi}_1^0})/m_{\tilde{\chi}_1^0}$:\\
~~a) $1.0 < \delta m_{\tilde{b}\tilde{\chi}_1^0}$~(red/dark),\\
~~b) $0.2 < \delta m_{\tilde{b}\tilde{\chi}_1^0} < 1.0$~(pink/grey),\\
~~c) $0.1 < \delta m_{\tilde{b}\tilde{\chi}_1^0} < 0.2$~(light-green/very light grey),\\
~~d) $\delta m_{\tilde{b}\tilde{\chi}_1^0} < 0.1$~(dark-green/light grey).

\begin{figure}[htbp]
\centerline{
\includegraphics[width=\linewidth]{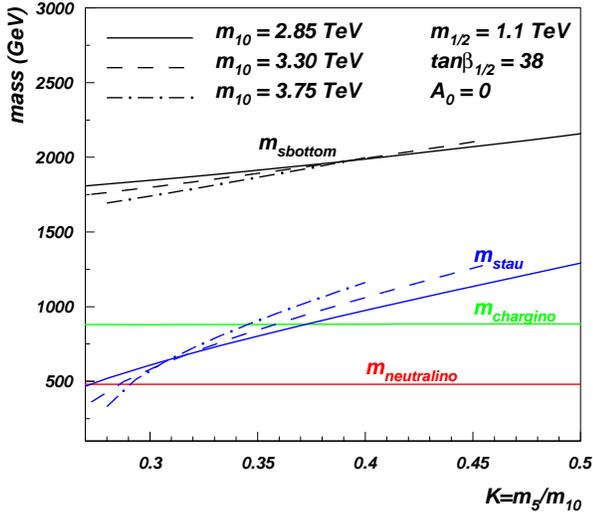}}
\caption{\label{fig:scan2}Sbottom and stau masses versus
$K=m_5/m_{10}$ for mSFNU benchmark points from \cite{Profumo:2003ema}
mass spectrum which was calculated using ISAJET v7.79.}
\end{figure}

One can see that the SBC scenario cannot be realized in the mSFNU scenario adopting a
more realistic approach which we used in our study. As indicated in
Figure~\ref{fig:scan1}, the neutralino-sbottom mass split $\delta
m_{\tilde{b}\tilde{\chi}_1^0}<0.2$ which is required by the SBC scenario, defines a
non-universal condition $m_Q<0.5 m_U$ which eventually cannot be realized in the
mSFNU scenario with  $m_Q= m_U$. We have also found that in the mSFNU region for the
SBC scenario from \cite{Pallis:2003aw,Profumo:2003ema}, actually, the stau co-annihilation takes
place when the ISAJET approach is used for the mass spectrum calculation. We
illustrate this in Figure~\ref{fig:scan2} which presents sbottom and stau masses
versus $K=m_5/m_{10}$ for the mSFNU benchmark points from the mass spectra
in~\cite{Profumo:2003ema} for which  was calculated using ISAJET v7.79. One can see
that in the nSFNU case the stau-neutralino mass split becomes small for sufficiently
small $K$, providing a stau-coannihilation mechanism of CDM suppression while sbottom
is quite far from being NLSP.

\begin{figure}[htbp]
\centerline{
\includegraphics[width=\linewidth]{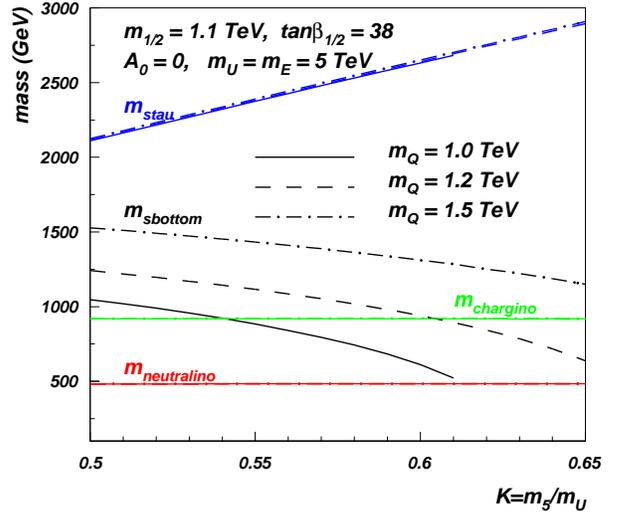}
}
\caption{
\label{fig:scan3}
Sbottom and stau masses versus
$K=m_5/m_U$ for  mSFNU extended by $m_Q$ as an additional
independent parameter which provides the SBC scenario}
\end{figure}

We have also found that if one departs from the mSFNU scenario and consider $m_Q$ as
an additional  independent parameter as we did in our scan, there are  solutions for
the SBC scenario. For sufficiently small values of $m_Q<0.5m_U$ as shown in
Figure~\ref{fig:scan3} the bottom squark becomes NLSP with a mass quite close to the
neutralino mass which provides the sbottom-neutralino co-annihilation suppression of
CDM. Hereafter we refer to this scenario as  next-to-minimal sfermion mass
non-universality (nmSFNU).

\begin{figure}
\includegraphics[width=\linewidth]{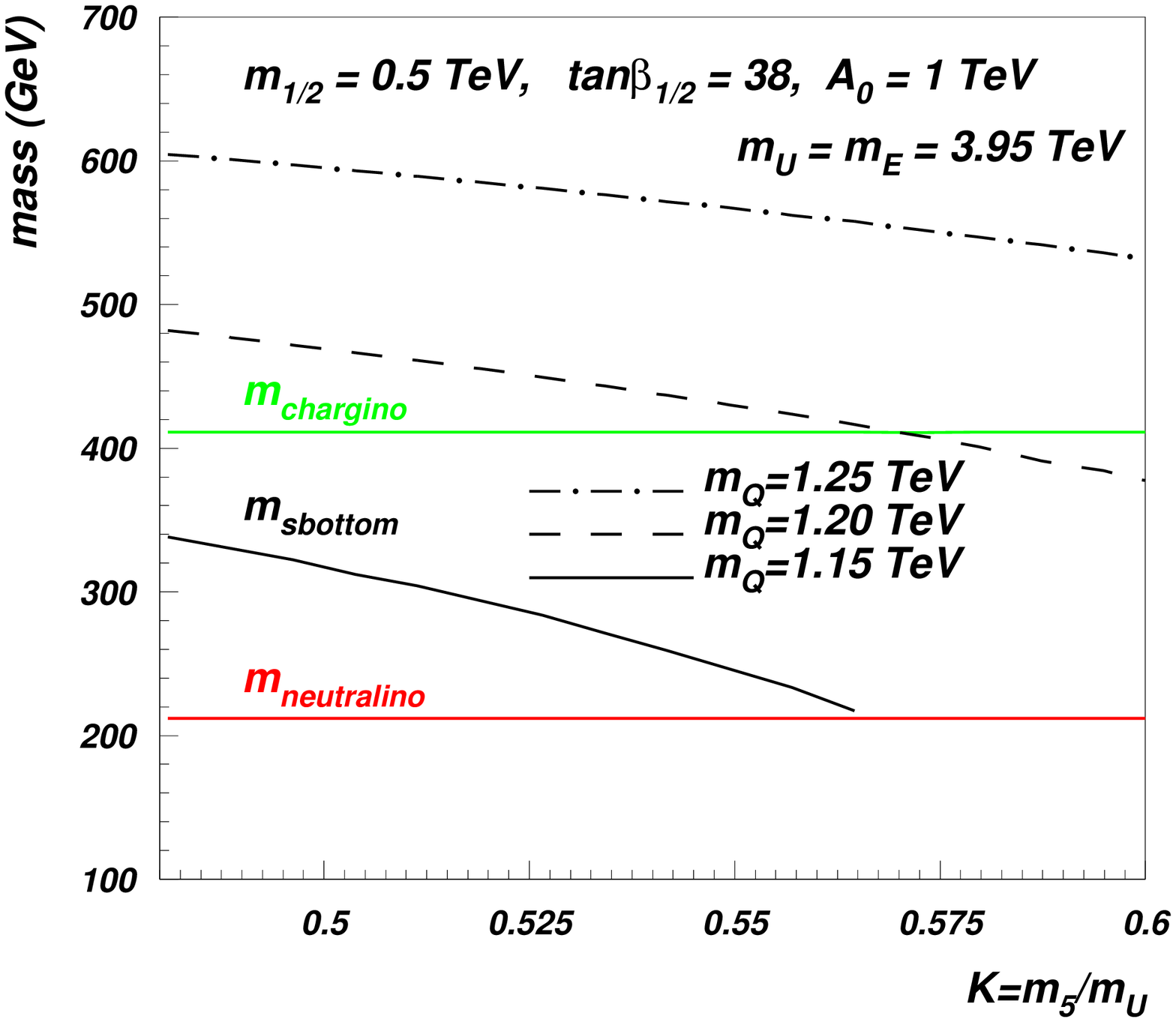}\\
\includegraphics[width=\linewidth]{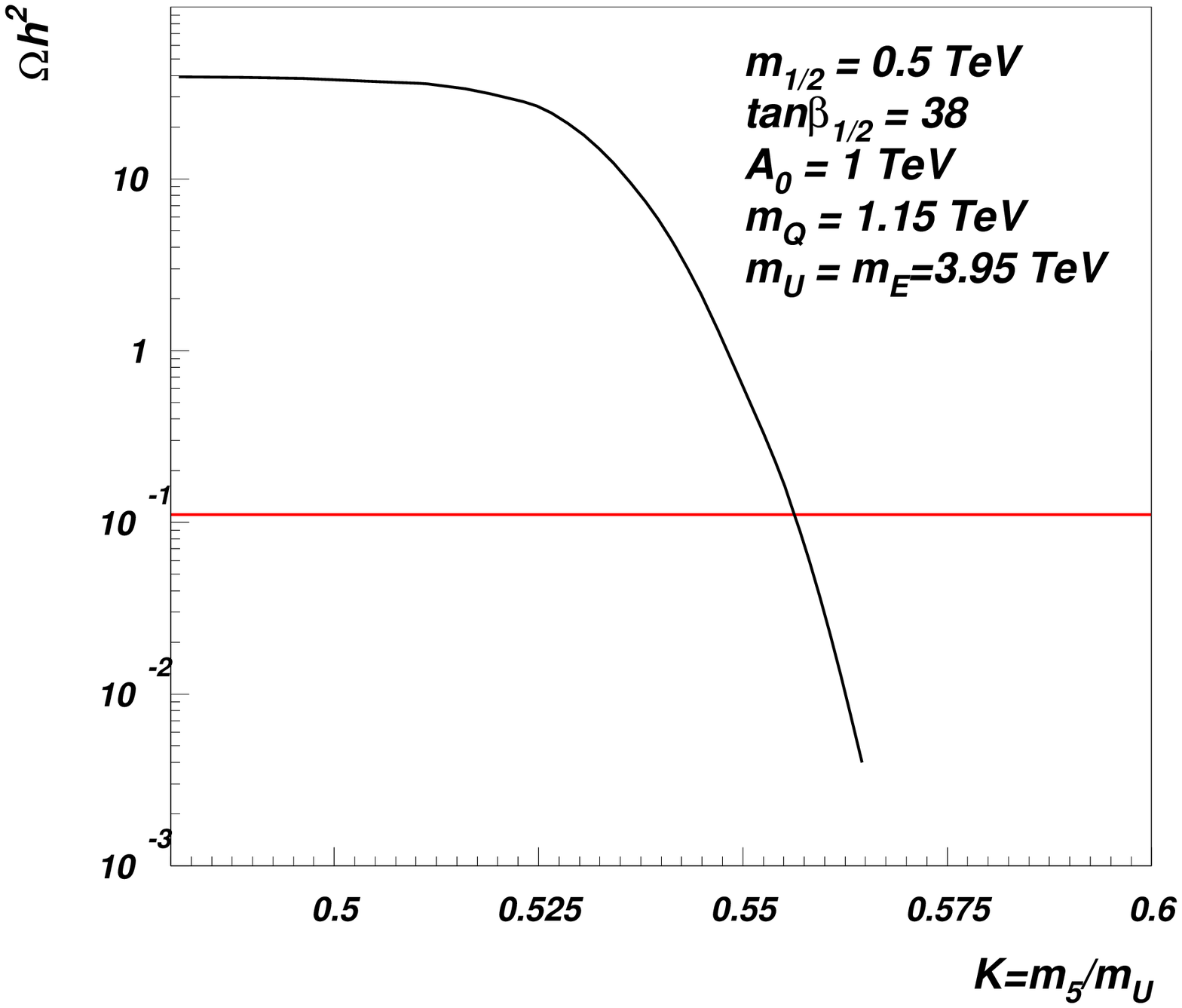}
\caption{\label{fig:scan4}
Sample point for nmSFNU
scenario which is relevant for 500\,GeV ILC:
$m_{1/2}=0.5$\,TeV,
$\tan\beta=38$, $A_0=1$\,TeV and  $m_U=m_L=3.95$\,TeV.
For sufficiently small $ m_Q$ parameter (of the order of 1\,TeV)
the SBC scenario takes place.
Top: sbottom mass versus $K=m_5/m_U$;
Bottom: $\Omega h^2$
versus $K=m_5/m_U$ which satisfies the $WMAP$ constraints
when $K \approx 0.555$ and $\delta m_{\tilde{b}\tilde\chi_1^0}\simeq 0.1$.}
\end{figure}

Finally, in Figure~\ref{fig:scan4} (top) we present a mass spectra for the nmSFNU
scenario for the benchmark point ($m_{1/2}=0.5$\,TeV, \mbox{$\tan\beta=38$},
$A_0=1$\,TeV and  $m_U=m_L=3.95$\,TeV) relevant to the 500\,GeV ILC which will be
able to produce sbottom with mass below 250 GeV with about 10\% relative mass split
with neutralino. For a sufficiently small $ m_Q$ parameter (of the order of 1\,TeV)
the SBC scenario takes place. In Figure~\ref{fig:scan4} (bottom) we present the
respective behaviour of $\Omega h^2$ versus $K=m_5/m_U$ which reaches $WMAP$
constraints when $\delta m_{\tilde{b}\tilde\chi_1^0}\simeq 0.1$.

We would like to stress that deviations from mSMNU which, as we have discussed,
provide a sbottom co-annihilation scenario would be an indicator of non-trivial
D-term contributions to soft scalar masses, which in its turn could give a hint about
the gauge group content at very high energies as noticed in~\cite{Kolda:1995iw}. As
we will show below, the ILC has a promising potential to cover parameter space
motivated by the SBC scenario thus shedding light on the gauge group properties at
the GUT scale.

\section{Sbottom production at the ILC}

\subsection{The ILC and the SiD Detector Concept}

\begin{figure}
\centerline{\includegraphics[width=7cm]{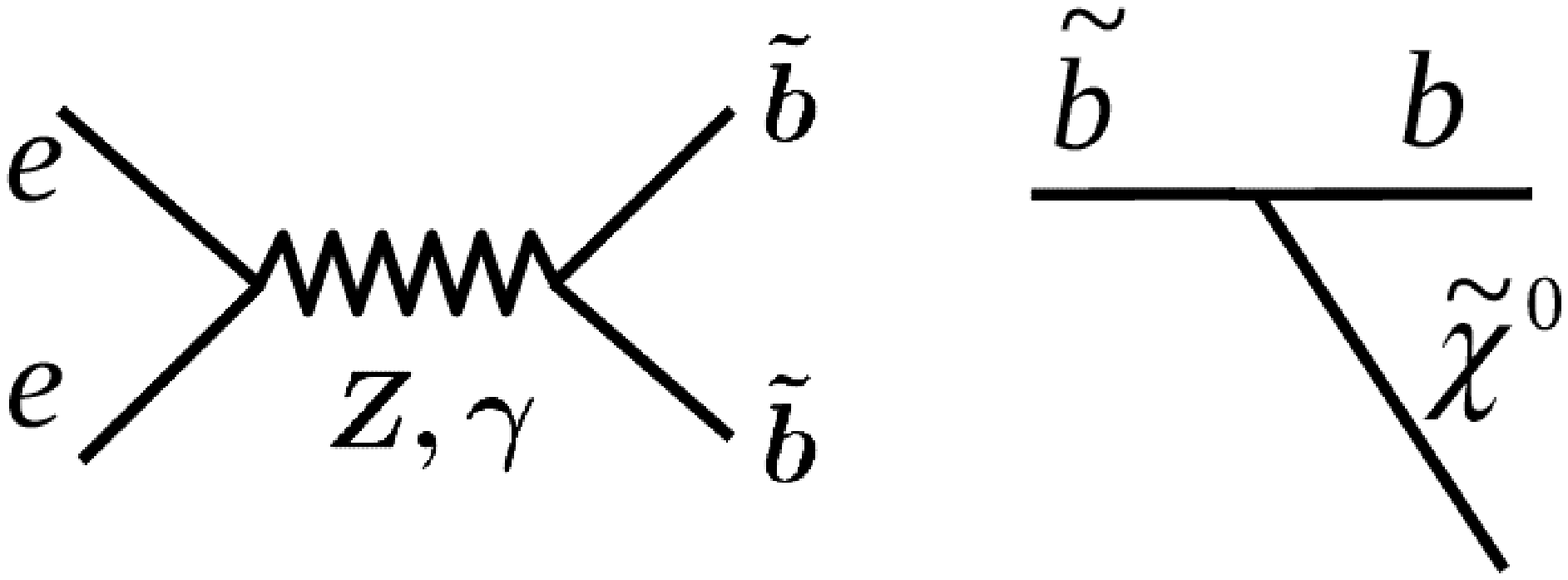}}
\caption{Feynman diagrams for sbottom pair production (left)
followed by sbottom decay into  $\tilde\chi_1^0$ and b-quark (right)}
\label{fig:diag}
\end{figure}

The ILC is a future electron-positron collider designed to collide particles at the
centre of mass energy of 500\,GeV. At the ILC sbottom pairs can be produced via
s-channel photon or $Z$-boson exchange followed by sbottom decay into
$\tilde\chi_1^0$ and b-quark, see Figure~\ref{fig:diag}. In the scenario described
above such events will result in two soft b-jets, which are generally not acoplanar
nor acolinear, and a missing energy.

\begin{figure*}
\centerline{\includegraphics[width=\linewidth]{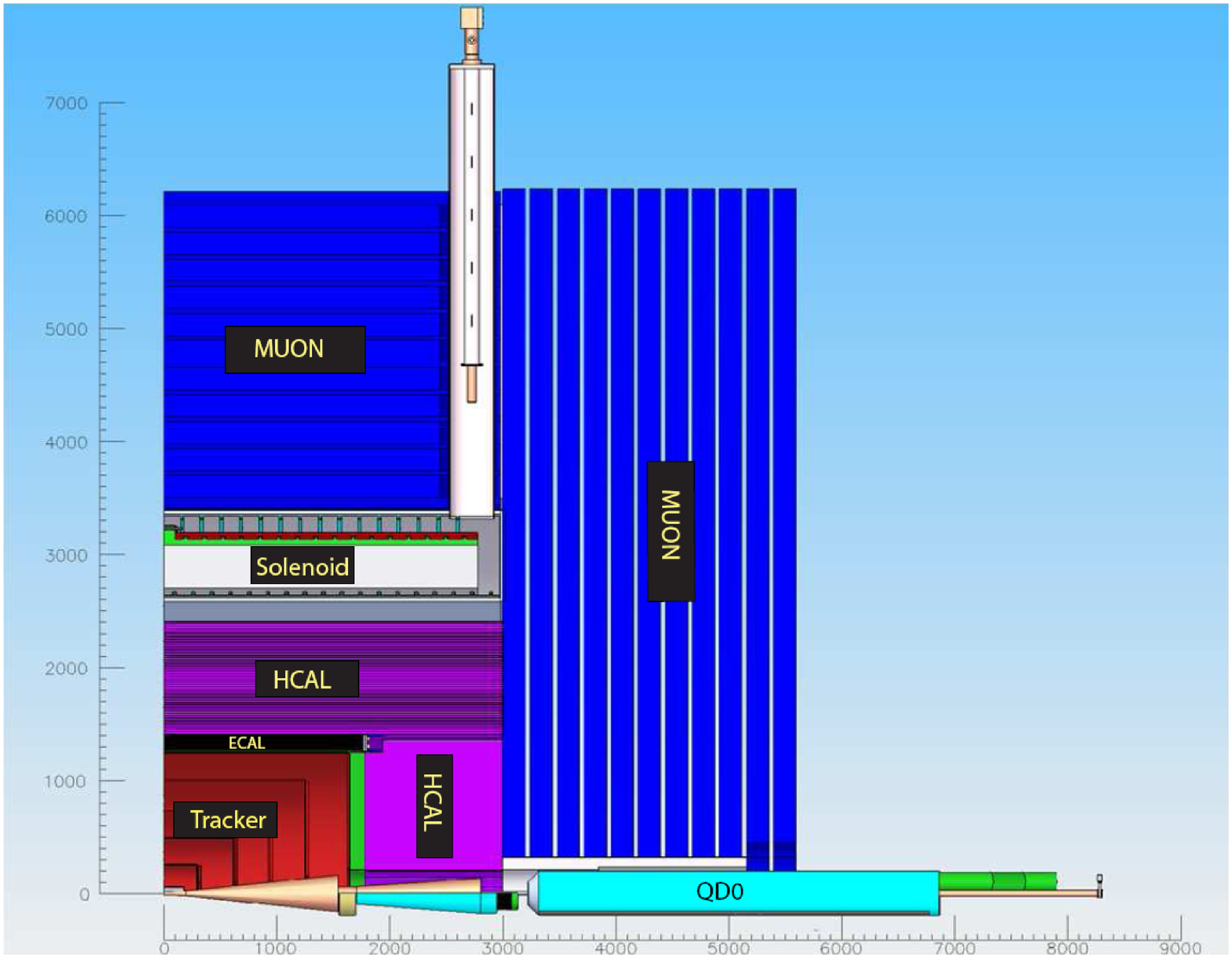}}
\caption{A plane view of a quadrant of SiD. Dimensions are in mm.}
\label{fig:SiD}
\end{figure*}

The sensitivity to the sbottom production was studied in the framework of the Silicon Detector (SiD) concept using full detector simulation and event reconstruction. SiD is a detector concept~\cite{SiD:LoI} designed for precision measurements of a wide range of possible new phenomena at the ILC. It is based on a silicon pixel vertex detector, silicon tracking, silicon-tungsten electromagnetic calorimetry, and a highly segmented hadronic calorimetry. Particle Flow Algorithm (PFA) approach~\cite{SiD:LoI} is an important strategy driving the basic philosophy and layout of the detector. SiD also incorporates a 5\,T solenoid, iron flux return and a muon identification system. A schematic view of SiD quadrant is shown in Figure~\ref{fig:SiD}.

\subsection{Signal and Background Rates and Properties}

In this study the CalcHEP package~\cite{Pukhov:2004ca}  was employed to generate SUSY
signal events. The events were converted to the Les Houches format~\cite{LesHouches},
passed to Pythia~\cite{pythia:man64} for their fragmentation and particle decays and
consequently to the SiD full simulation and reconstruction chain~\cite{SiD:LoI}.
Several points in the MSSM parameter space were chosen close to the kinematic limit
in order to investigate the discovery potential. These points correspond to various
masses of the bottom squark and neutralino: (230, 210); (240, 210); (230,220);
(240,220); (220,210), all in GeV. For each signal point 200k events were generated,
also accounting for the ISR, FSR and Beamstrahlung (BS) processes.

\begin{figure}
\centerline{\includegraphics[width=\linewidth]{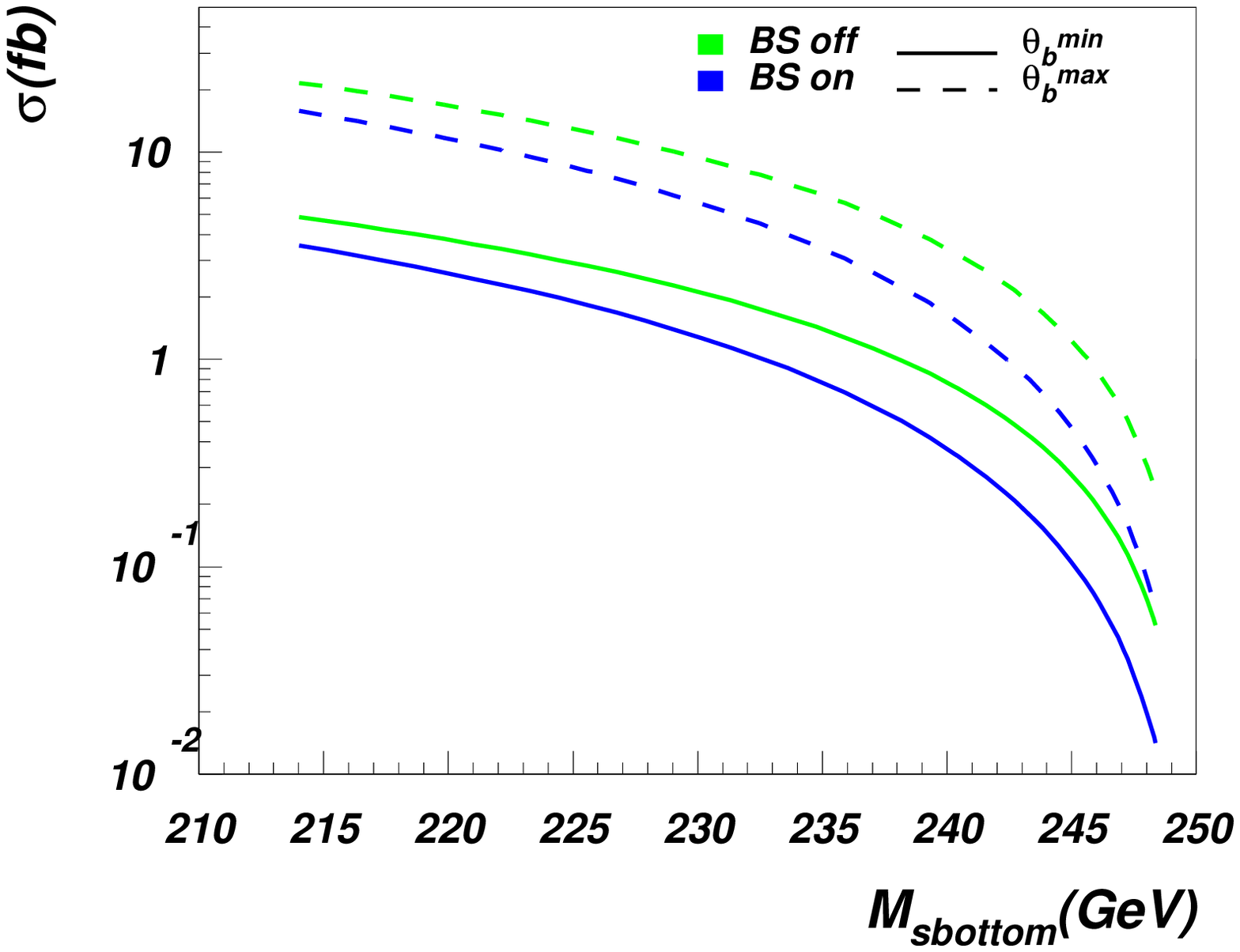}}
\caption{ Sbottom pair production cross section at ILC at \mbox{$\sqrt{s} =
500$\,GeV}:  solid line corresponds to the case of the minimal cross section which
takes place at $cos\theta_b^{min}\simeq 0.35$ while the dashed line corresponds to
the case of the maximal  cross section which takes place at $cos\theta_b^{max} = 1$.
The line (light/dark grey) colors present the effect of Beamstrahlung radiation (on/off). }
\label{fig:XS}
\end{figure}

The cross section of sbottom pair production depends strongly on how close the
sbottom mass is to the ILC kinematical limit, see Figure~\ref{fig:XS} for
\mbox{$\sqrt{s} = 500$\,GeV}, diving quickly down for masses approaching 250\,GeV.
The mass of the lightest bottom squark mass, $m_{\tilde{b}}$, and the mixing angle of
the lightest and heaviest bottom squarks, $\theta_b$, completely determine the cross
section of pair production of lightest bottom squarks. For the fixed $m_{\tilde{b}}$
mass, the cross section can vary by factor of about 5, depending on the value of
$\theta_b$. The solid line in Figure~\ref{fig:XS} corresponds to the case of the
minimal cross section, which takes place at $cos\theta_b^{min}\simeq 0.35$, while the
dashed line corresponds to the case of the maximal cross section which takes place at
$cos\theta_b^{max} = 1$. The line grey levels (light/dark) in Figure~\ref{fig:XS} present
the effect of Beamstrahlung radiation (on/off), which decreases the cross section by
about 20\%. The signal cross section is about $4-1$\,fb for $215-230$\,GeV bottom
squark mass for $\theta_b^{min}$ (solid dark line, Figure~\ref{fig:XS}) and about
factor of 5 larger for  $\theta_b^{max}$ (dashed dark line, Figure~\ref{fig:XS}). In
our analysis we use signal cross section  $\theta_b^{min}$, therefore conservatively
estimating the ILC potential to probe the SBC scenario.

\begin{figure}
\includegraphics[width=\linewidth]{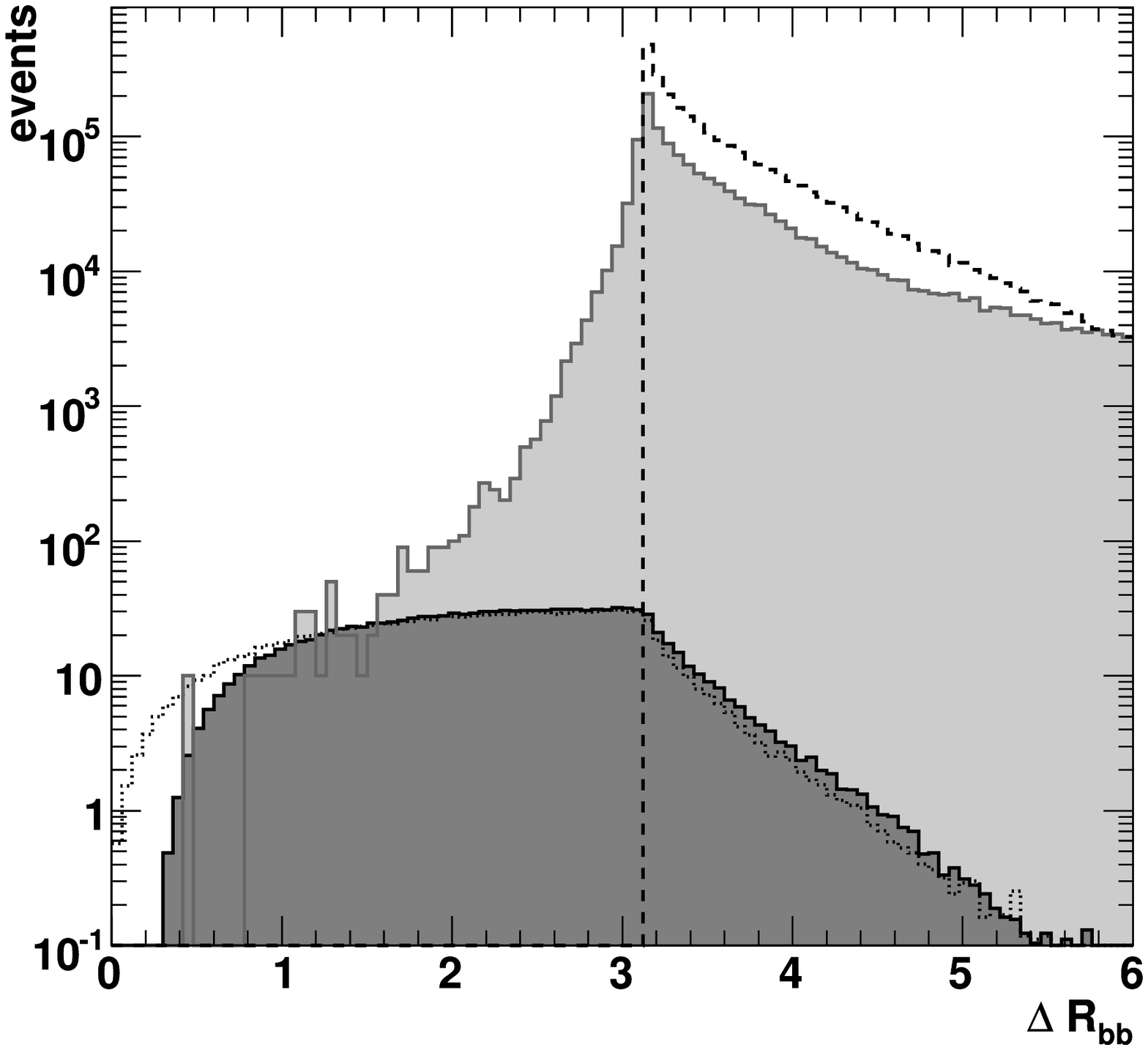}
\caption{
$\Delta R_{bb}=\sqrt{\Delta_\phi^2+\Delta_\eta^2}$ parton-level distribution for
$e^+e^-\to\tb_1\bar{\tb_1}\to b \bar{b}\tilde{\chi}_1^0\tilde{\chi}_1^0$ signal (230,210)
(empty histogram with dotted line border)
and $e^+e^-\to b\bar{b}$ background (empty histogram with dashed line border)
processes compared to the
full detector simulation level for signal (light grey shaded histogram)
and background (light grey shaded histogram). } \label{fig:dr_part}
\end{figure}

While the signal is at a few fb level, the main background, which originates
essentially from quasi-real photons emitted by electron and positron at very low
scattering angles, $\gamma\gamma\to b\bar{b}$, is about 4.5\,pb at 500\,GeV ILC.
Another significant background of similar topological properties is the di-jet
background from $e^{+}e^{-} \to b\bar{b}$ which has about 2.7\,pb production cross
section as well as $e^+e^- \to q\bar{q}$, where $q$ represents other quarks
misidentified as $b$. One of the powerful variables, which could suppress such
dominant backgrounds, is the b-jet separation variable $\Delta
R_{bb}=\sqrt{\Delta_\phi^2+\Delta_\eta^2}$, where $\Delta_\phi$ and $\Delta_\eta$ is
the difference of jet azimuthal angles and rapidities, respectively. In the case of
backgrounds with two back-to-back b-jets $\Delta R_{bb}>\pi$ while in the case of
signal, which has also two neutralinos in the final state together with two b-jets,
$\Delta R_{bb}$ can be well below $\pi$. This is illustrated in
Figure~\ref{fig:dr_part}.

Though the parton level $\Delta R_{bb}$ distribution for signal versus background looks very promising
and suggests to eliminate the $e^+e^- \to b\bar{b}$ and $\gamma\gamma\to b\bar{b}$ backgrounds completely by setting $\Delta R_{bb}>\pi$ cut,
the situation is not that
optimistic at the full detector simulation level: $\Delta R_{bb}$ distribution from $\gamma\gamma\to b\bar{b}$ and $e^{+}e^{-} \to b\bar{b}$
processes can "leak" into $\Delta R_{bb}<\pi$ region due to detector resolution and event reconstruction effects. The realistic
$\Delta R_{bb}$ distribution after the full detector simulation is shown as filled histograms in Figure~\ref{fig:dr_part}.

Complete sets of unbiased Standard Model processes at 500\,GeV were generated using the Whizard Monte Carlo program~\cite{whizard} with a total statistics of about 7M events. All 0, 2, 4 and 6 fermion final states were generated, see~\cite{SiD:LoI, SiD:MCweb} for details. PYTHIA~\cite{pythia:man64} was used for final state QED and QCD parton showering, fragmentation and decay to provide final-state observable particles. Included in this sample are backgrounds arising from interactions between virtual and beamstrahlung photons. Both signal and background samples are considered un-polarised. The detector simulation is based on Geant~4 toolkit~\cite{geant4,geant4dev}. A thin layer of Linear Collider specific code, SLIC~\cite{slic}, provides access to the Monte Carlo events, the detector geometry and the output of the detector hits.



\subsection{Analysis and Results}

\begin{figure}
\centerline{\includegraphics[width=\linewidth]{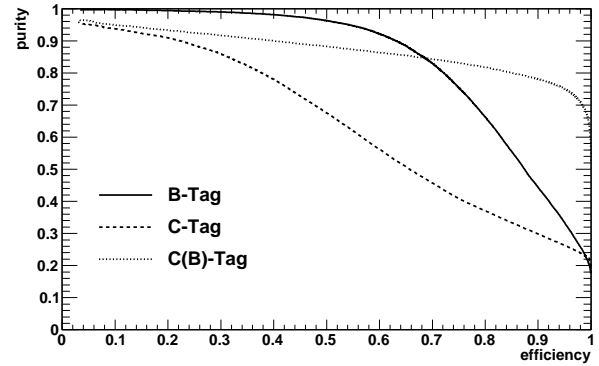}}
\caption{Flavour tagging performance of the LCFI package in terms of purity vs. efficiency optimised for the SiD detector concept and evaluated for di-jet events $e^+e^- \rightarrow q\bar{q}$. Three tags provided are shown: b-tag, c-tag and c-tag on the b-background only (c(b)-tag).}
\label{fig:purity_efficiency}
\end{figure}

The vertexing package used for the jet flavour tagging in this analysis was developed by the LCFI collaboration~\cite{LCFI:Package}. The main algorithm of the package, the topological vertex finder ZVTOP, reconstructs vertices in arbitrary multi-prong topologies. It combines relevant variables in neural networks which are then separately trained on samples of b-, c- and light quarks. The most
discriminating variables are the corrected vertex mass, the joint probability, the secondary vertex probability, the impact parameter significance of the most significant track and number of vertices in the event. The joint probability is defined as the probability for all tracks in a jet to be compatible with hypothesis that they originate at the primary vertex. The performance of the LCFI package optimised for the SiD detector is shown in Figure~\ref{fig:purity_efficiency} which shows the flavour tagging purity as a function of efficiency for the provided three tags.

The SiD specific optimisation was performed by building new neural networks and training them using di-jet samples which passed through the full SiD simulation and reconstruction.

\begin{figure}[htbp]
\centerline{\includegraphics[width=\linewidth]{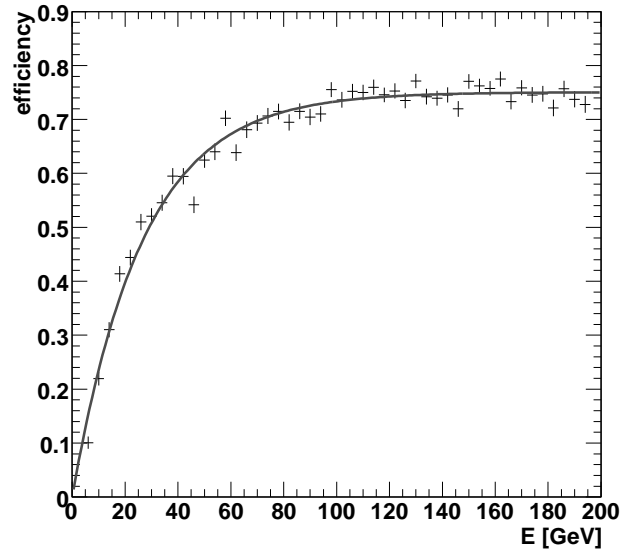}}
\caption{Jet b-tagging efficiency as a function of the jet energy for b-jets from di-jet events $e^+e^- \rightarrow q\bar{q}$.}
\label{fig:btag_efficiency}
\end{figure}

Observation of the sbottom production is challenging due to a very low energy of
b-jets, below 20-30\,GeV, which pushes the jet clustering and tagging algorithms to
their limits. Figure~\ref{fig:btag_efficiency} illustrates that while the b-tagging
efficiency is about 75\% above 60~GeV, it is falling steeply for lower jet energies.
The energy of b-quarks is determined by the mass difference between $m_{\tilde{b}}$
and $m_{\tilde{\chi}_1^0}$.

\begin{figure*}[htbp]
\centerline{\includegraphics[width=\linewidth]{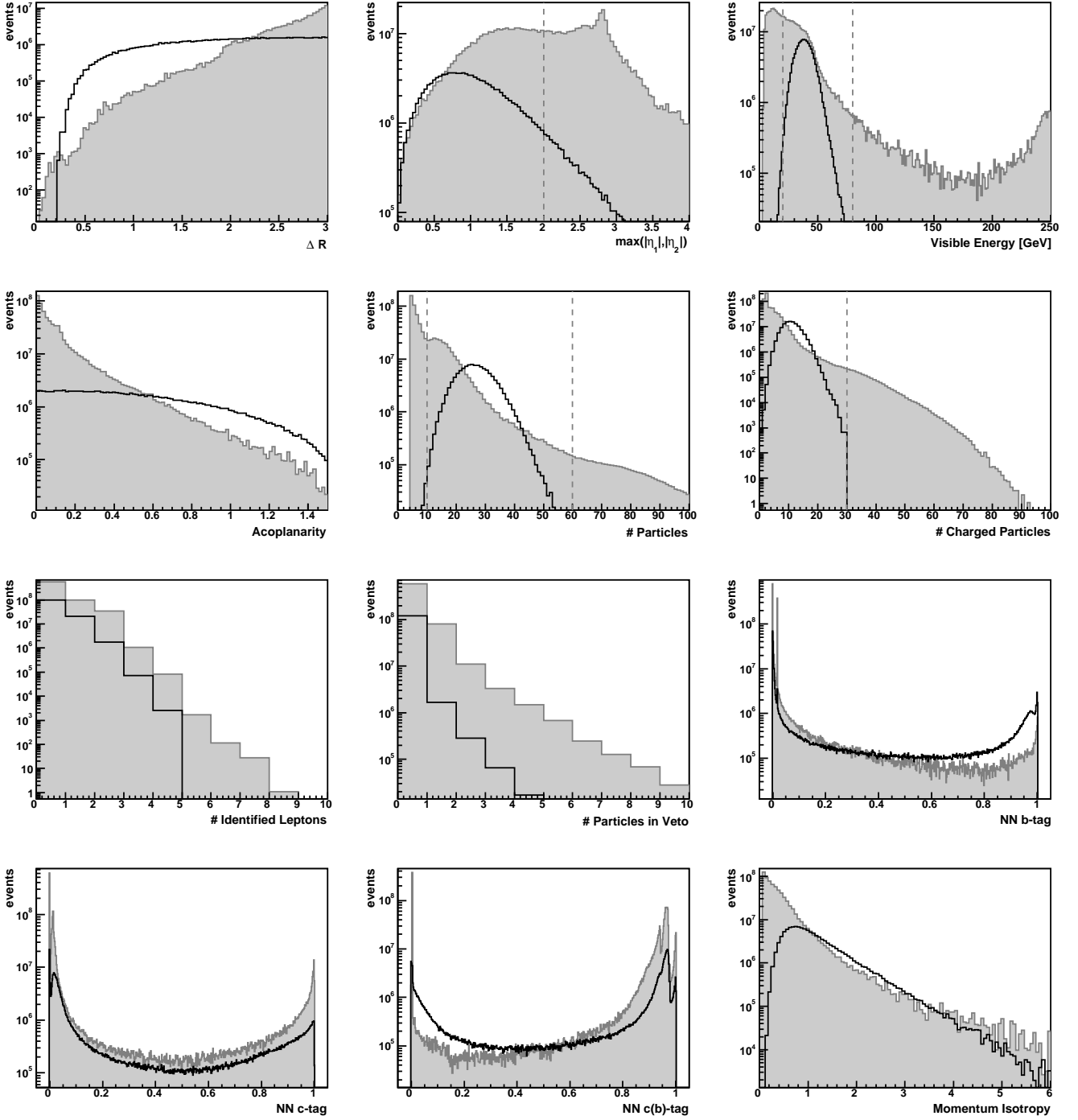}}
\caption{Neural net input variables. Signal with $m_{\tilde{b}} = 230\,$GeV and $m_{\tilde{\chi}_1^0} = 210\,$GeV (open histogram) is shown together with the inclusive SM background (filled histogram). The signal was scaled up by a factor of 100000. The dashed lines indicate selection cuts applied before neural net training, see text for details. }
\vspace{1cm}
\label{fig:variables}
\end{figure*}

Jets were reconstructed by employing the Durham $k_T$ algorithm with $k_T^{min} = 10\,$GeV and limiting the maximum number of jets to two. Figure~\ref{fig:variables} shows distributions of several variables used in the analysis for the signal with $m_{\tilde{b}} = 230\,$GeV and $m_{\tilde{\chi}_1^0} = 210\,$GeV and compares them to the inclusive SM background. The signal was scaled up by a factor of $10^5$ for the presentation purposes. The variables are (from top-left): distance between jets $\Delta R$ in $\eta - \phi$ plane, maximum of jet rapidities $max(|\eta_1|,|\eta_2|)$, visible energy, acoplanarity, number of particles in event, number of charged particles, number of identified leptons, number of particles in the forward calorimeter used for veto, neural nets outputs of jets for b-tag, c-tag, c(b)-tag and the momentum isotropy.

Majority of the SM background, for the exception of the two-photon background, is suppressed by the $E_{visible} < 80\,$GeV
selection. The signal events are misbalanced in energy due to the neutralinos. The dominant two-photon and $e^+e^- \to q\bar{q}$ backgrounds produce
jets in a back-to-back topology motivating the \mbox{$\Delta R_{\eta\phi} < 3.0$}
selection. To further suppress the background $max(|\eta_1|,|\eta_2|) < 2.0$
is required, where $\eta_1$ and $\eta_2$ are jet pseudorapidities. Finally,
the total number of reconstructed particles in an event is required to be within $10 \leq
N_{particles} \leq 60$ for the mass point (230,210) and it is adjusted for
the other mass points individually.

An important ingredient in this analysis is an electromagnetic veto using the forward detector at very low polar angles, above $\theta = 10\,$mrad. This is used to suppress the two-photon background as well as other backgrounds with a large ISR contribution. The forward detector acceptance was estimated using a simple geometrical model. If a photon or electron with $E>300\,$MeV is detected within the acceptance the event veto is applied.

\begin{figure*}[htbp]
\centerline{
\includegraphics[width=0.48\linewidth]{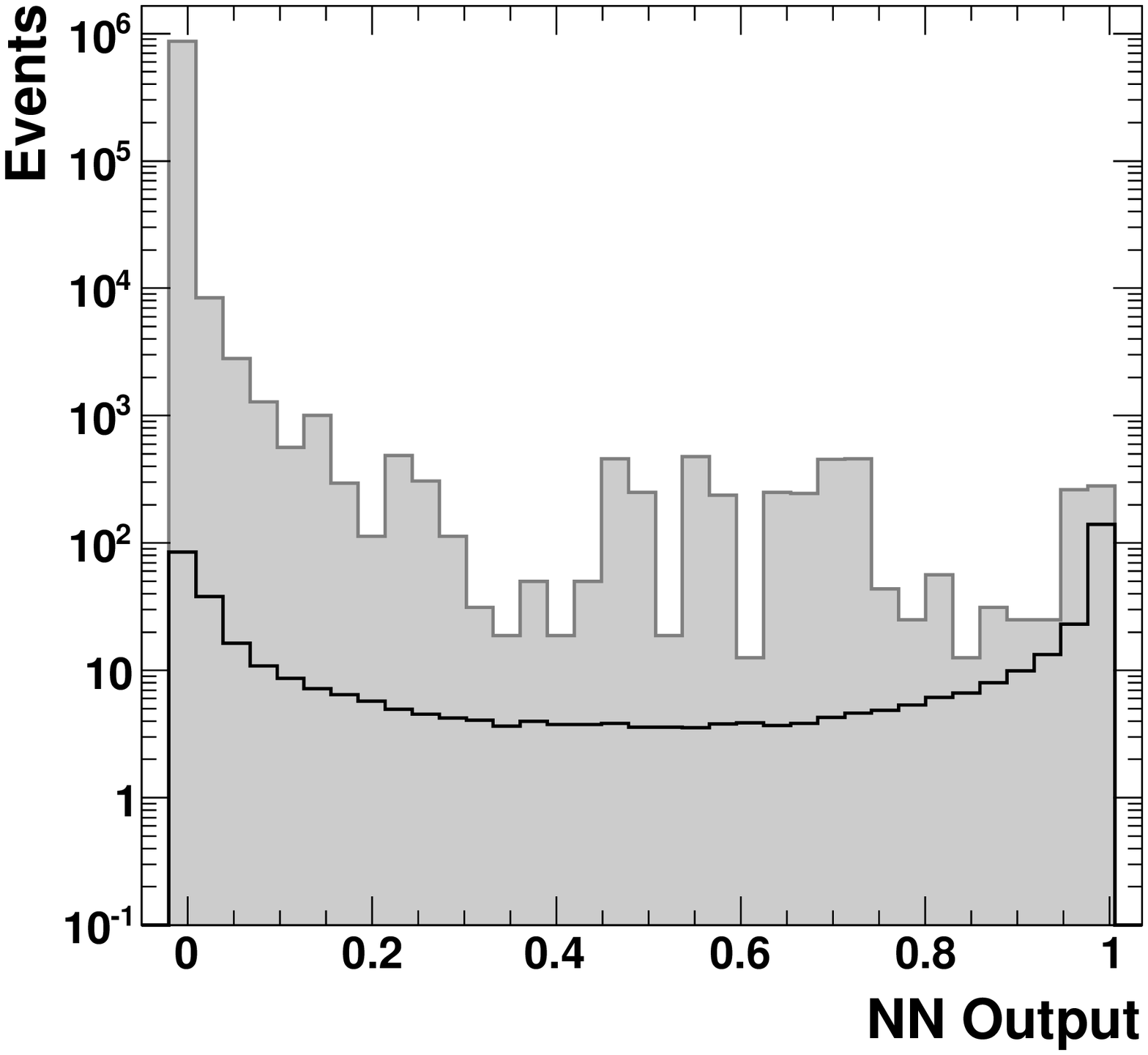}
\includegraphics[width=0.48\linewidth]{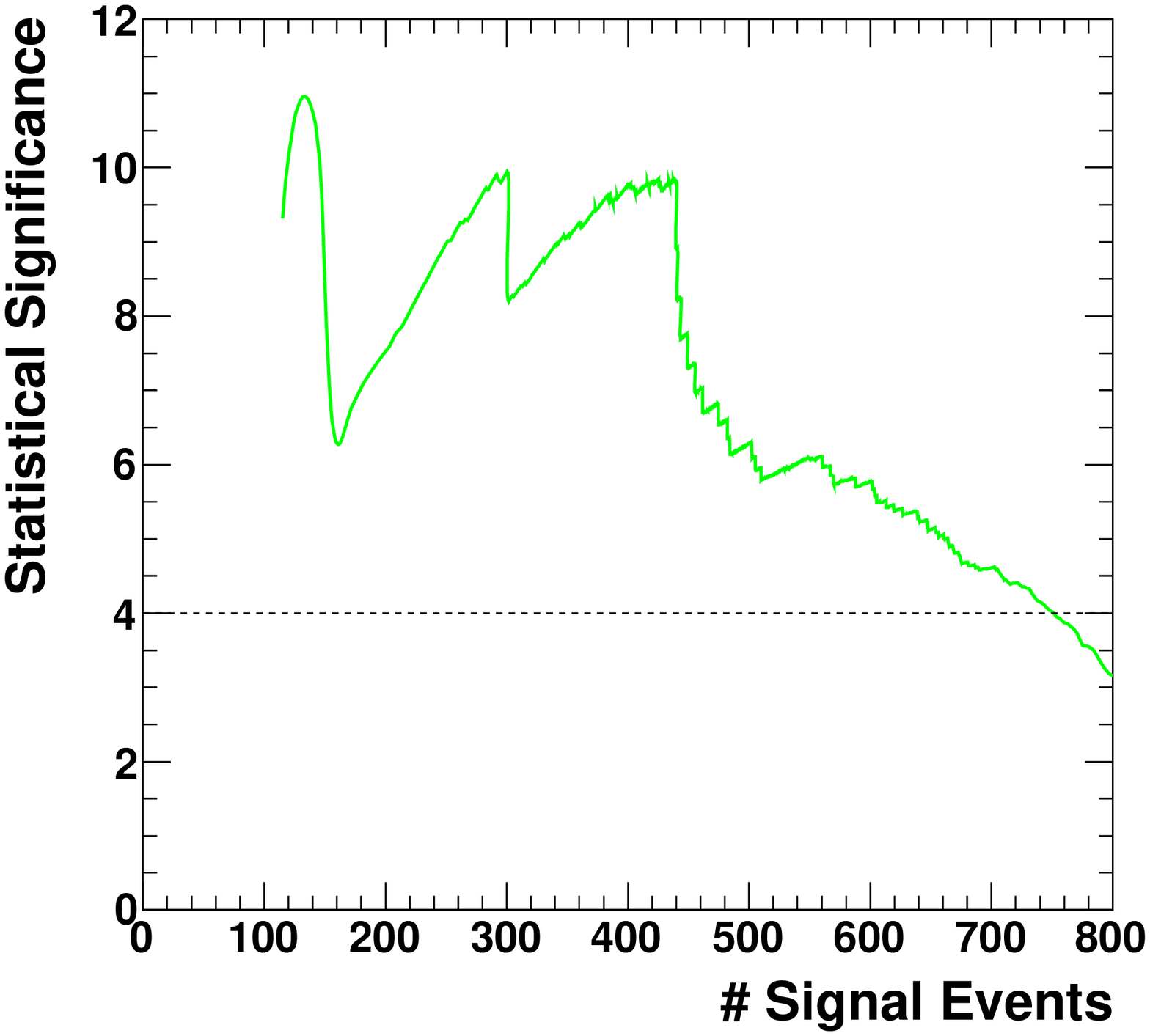}
}
\caption{Neural network output (left) is shown for events which passed the basic selections for signal with $m_{\tilde{b}} = 230\,$GeV and 
$m_{\tilde{\chi}_1^0} = 210\,$GeV (open histogram) and inclusive SM background (filled histogram). Statistical significance (right) is shown as a function of a number of 
selected signal events. Dashed line shows a limit of four standard deviations. Big variations are caused by SM events with large weights.}
\label{fig:confidence_level}
\end{figure*}

For the final event selection a FANN neural network~\cite{fann}  based on the above variables was defined and trained on independent signal and background samples.
The resulting neural net output is shown in the left plot of Figure~\ref{fig:confidence_level}. The result is interpreted in terms of signal significance calculated as $S/\sqrt{S+B}$, where $S$ is the number of selected signal events and $B$ the number of selected background events. The event numbers are normalised to the total luminosity of $1000~$fb$^{-1}$. Figure~\ref{fig:confidence_level} shows a distribution of $S/\sqrt{S+B}$ as a function of the number of selected signal events with each bin corresponding to a particular selection of the final cut on the neural net classifier output. Big variations are caused by the background events with large weights. The large weights were used for some SM processes with large cross sections.

\begin{figure}[htbp]
\centerline{\includegraphics[height=7cm]{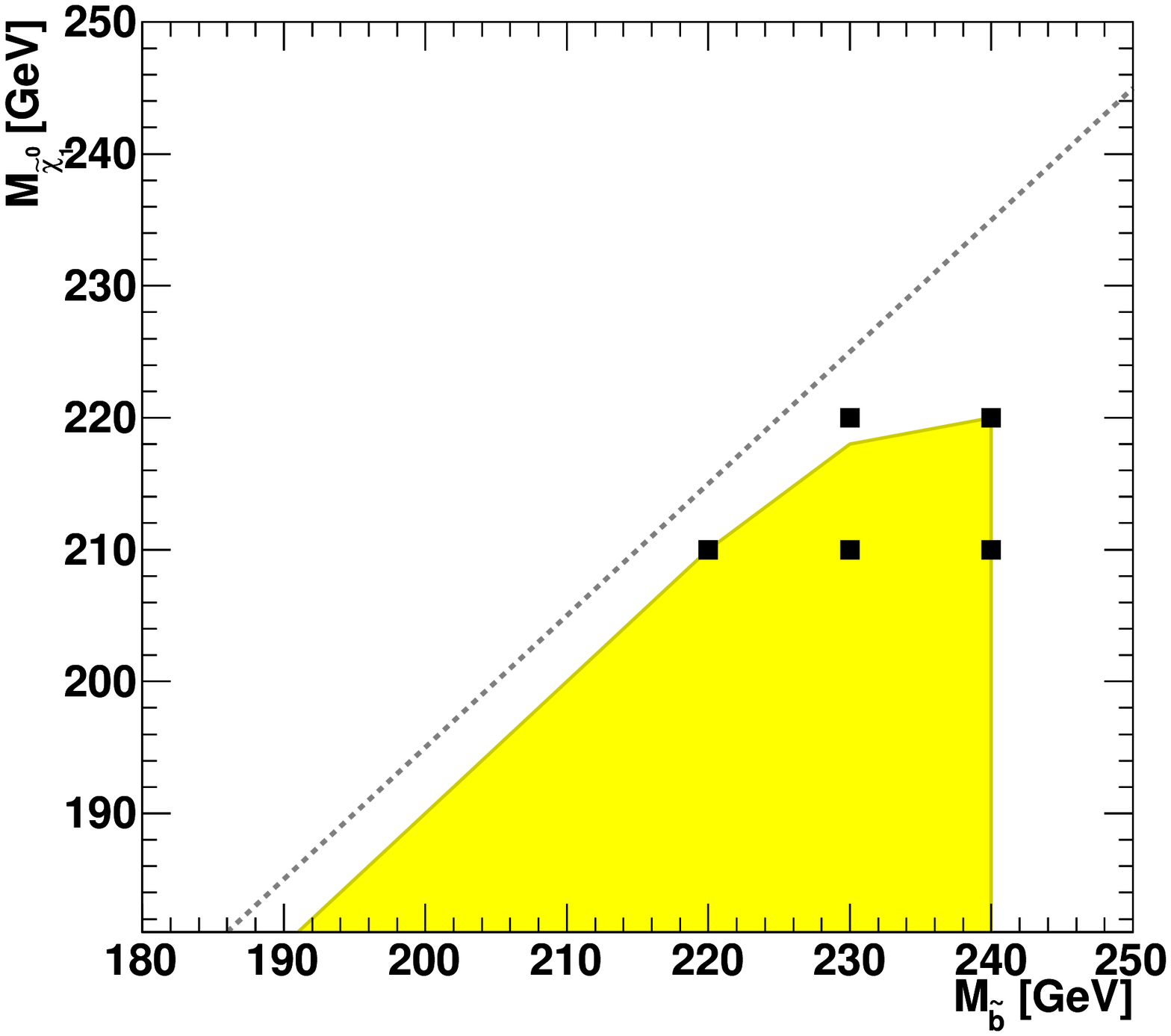}}
\caption{Selected mass points in the $m_{\tilde{b}}$ - $m_{\tilde{\chi}_1^0}$ plane.
The dashed line corresponds to the kinematic limit corresponding to a small mass
difference between $m_{\tilde{b}}$ and $m_{\tilde{\chi}_1^0}$. The filled area is an estimate of the 3$\sigma$ observation region.}
\label{fig:kinematic_plane}
\end{figure}

Based on the above the signal cross section statistical uncertainty was calculated to be equal to 15\,\% for the (230,210) mass point. The mass points (240, 210); (220,210); (240,220) all can be observed with 3\,$\sigma$ statistical significance and the mass point (230,220) with 2\,$\sigma$ statistical significance. A conservative estimate of the 3\,$\sigma$ observation region is shown in Figure~\ref{fig:kinematic_plane}.

\section{Summary}

We have performed analysis of ILC potential to probe a sbottom
co-annihilation (SBC) scenario. We have found that if the SBC scenario is
confirmed at the ILC, this would  be an indication of related non-minimal
sfermion non-universality scenario at the GUT scale reflecting non-trivial
D-term contributions to soft scalar masses. This would give a hint about
the gauge group content at very high energies.

The probe of SBC scenario is very challenging even at the ILC due to the
fact that background is higher than the signal by several orders of
magnitude. One should stress, that the realistic detector simulation and
reconstruction is crucial for this study due to the important role of
resolution effects.

Using neural network analysis with the optimised input variables we
conclude that the SBC scenario can be probed very close to the ILC
kinematic limits and, in the case of discovery, the sbottom production
cross section can be measured with a reasonable statistical precision. Such
a measurement, however, relies on a presence of the forward detector used
to veto two-photon background events and soft di-jet events with a dominant
ISR contribution.

The approximate reach of the 500\,GeV ILC extends to within ~10\,GeV to the
kinematic limit for the sbottom pair production and within 5\,GeV to the
kinematic limit of the sbottom decay to LSP and b-quark.

\acknowledgments{
We would like to thank the SiD software and benchmarking groups and LCFI collaboration,
in particular Jan Strube, Tim Barklow, Norman Graf and M\'{a}rija Kov\'{a}\v{c}evi\`{c} for assistance with sample processing and useful discussions.
A.B. is grateful for the hospitality of GGI Institute where the final version of the paper has been prepared.
}

\bibliographystyle{apsrev}
\newpage

\bibliography{sbottom}

\end{document}